\documentclass[aps,pre,twocolumn,superscriptaddress,amsmath,amssymb,longbibliography]{revtex4-1}

\usepackage{graphicx}
\usepackage{dcolumn}
\usepackage{bm}
\usepackage{hyperref}
\usepackage{comment}
\usepackage{color}
\usepackage{tabularx}
\newcolumntype{f}{X}
\newcolumntype{n}{>{\hsize=.9\hsize}X}
\newcolumntype{k}{>{\hsize=.75\hsize}X}
\newcolumntype{u}{>{\hsize=.5\hsize}X}
\newcolumntype{q}{>{\hsize=.25\hsize}X}
\begin{document}

\title{Active Spaghetti: Collective Organization in Cyanobacteria}

\author{Mixon K. Faluweki}
\thanks{equal contribution}
\affiliation{School of Science and Technology, Nottingham Trent University, Nottingham NG11 8NS, UK}
\affiliation{Malawi Institute of Technology, Malawi University of Science and Technology, S150 Road, Thyolo 310105, Malawi}
\author{Jan Cammann}
\thanks{equal contribution}
\affiliation{Interdisciplinary Centre for Mathematical Modelling and Department of Mathematical Sciences, Loughborough University, Loughborough, Leicestershire LE11 3TU, United Kingdom}
\author{Marco G. Mazza}
\email{E-mail: m.g.mazza@lboro.ac.uk}
\affiliation{Interdisciplinary Centre for Mathematical Modelling and Department of Mathematical Sciences, Loughborough University, Loughborough, Leicestershire LE11 3TU, United Kingdom}
\affiliation{Max Planck Institute for Dynamics and Self-Organization (MPIDS), Am Fa{\ss}berg 17, 37077 G\"{o}ttingen, Germany}
\author{Lucas Goehring}
\email{E-mail: lucas.goehring@ntu.ac.uk}
\affiliation{School of Science and Technology, Nottingham Trent University, Nottingham NG11 8NS, UK}

\date{\today}

\begin{abstract}
Filamentous cyanobacteria can show fascinating examples of nonequilibrium self-organization, which however are not well-understood from a physical perspective. We investigate the motility and collective organization of colonies of these simple multicellular lifeforms.  As their area density increases, linear chains of cells gliding on a substrate show a transition from an isotropic distribution to bundles of filaments arranged in a reticulate pattern. Based on our experimental observations of individual behavior and pairwise interactions, we introduce a nonreciprocal model accounting for the filaments' large aspect ratio, fluctuations in curvature, motility, and nematic interactions.  This minimal model of active filaments recapitulates the observations, and rationalizes the appearance of a characteristic lengthscale in the system, based on the P\'eclet number of the cyanobacteria filaments. 
\end{abstract}

\maketitle

Collective organization is a defining feature of living matter. It has received vivid attention~\cite{vicsek1995,toner1995,simha02,ramaswami2010,marchetti2013,elgeti2015,fruchart2021,Shi2018} for its applications in the life sciences~\cite{needleman2017,doostmohammadi2018}, and as an example of how nonequilibrium forces can drive flows of matter and energy~\cite{battle2016,cammann2021}. The first seminal studies of active matter treated the motion of point-like particles~\cite{vicsek1995,toner1995,ramaswami2010}.  Non-reciprocal interactions between even such simple objects, with a single orientation, allow access to states impossible in equilibrium systems \cite{fruchart2021}, and rod-like motile particles extend the range of such emergent behavior~\cite{Shi2018}.   Long, flexible filaments, whose orientation varies along their length, offer opportunities to study different classes of active matter \cite{liverpool2001, isele2015, jiang2014, duman2018, bianco2018, anand2018, joshi2019,fily2020, winkler2020, peterson2020,du2022,abbaspour2021}. With many possible interaction points per filament, correlations can spread over long distances, opening the door to novel behavior \cite{denk2016,prathyusha2018,winkler2020,sumino2012,tamulonis2014,chelakkot2021,sciortino2021}, whose complete understanding remains lacking.  

\begin{figure*}
    \centering
    \includegraphics[width=1\linewidth]{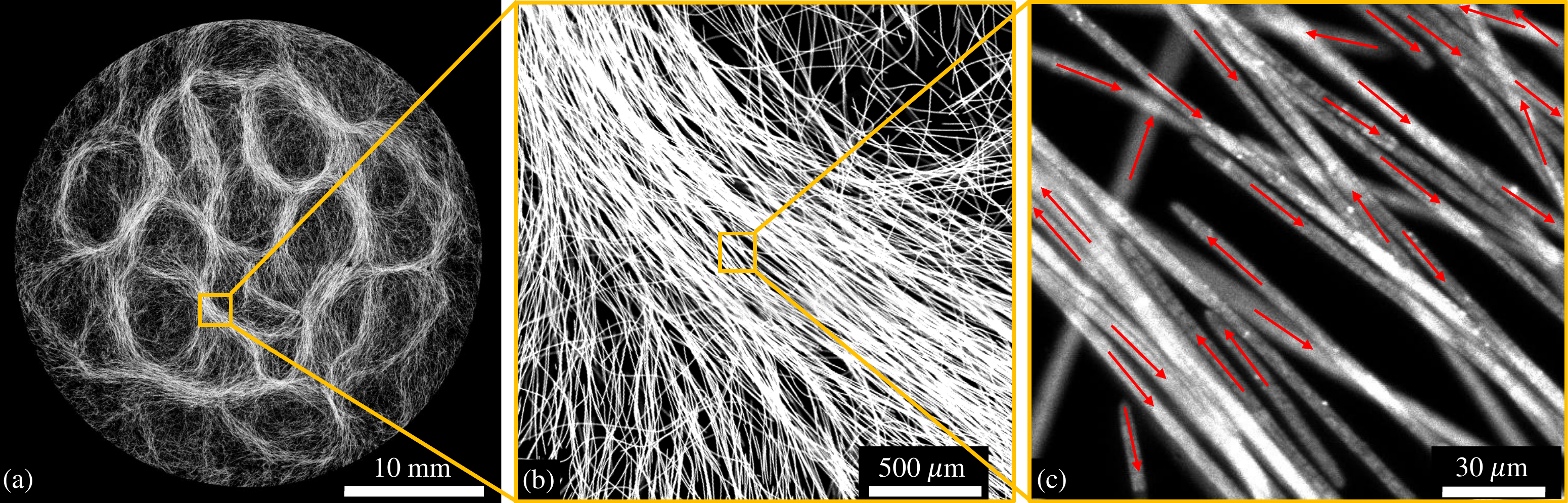}
    \caption{A colony of \emph{O. lutea} at density $\rho = 53$\,mm$^{-2}$ shows (a) a reticulate pattern, with (b) the local alignment of filaments within bundles, and (c) filament motion (arrows) that is predominantly parallel or anti-parallel to neighbors.}
    \label{fig:motivation_fig}
\end{figure*}

An important example of active matter, cyanobacteria are among the Earth's most abundant and ancient organisms~\cite{bivzic2020,Sumner1997}. They evolved the original mechanisms of photosynthesis and perform nearly all nitrogen fixation in marine environments~\cite{Kasting2002,Capone1997}. Filamentous cyanobacteria also straddle the boundary between single and multicellular organisms; they grow into long chains of cells through `filamentation', perhaps the oldest form of multicellularity~\cite{Schirrmeister2011,mizuno2022}. Many species live on surfaces, including stromatolites~\cite{Schirrmeister2011,Reid2000}, and move by gliding~\cite{halfen1970,hoiczyk2000,read2007}.   Colonies can develop complex structures of closely-bundled filaments, such as reticulate patterns (Fig.~\ref{fig:motivation_fig}), over hours or days~\cite{Shepard2010UndirectedMats,tamulonis2014,cuadrado2018}.  Cell density is thought to be a trigger of such pattern formation~\cite{Shepard2010UndirectedMats,tamulonis2014}, but this link has never been conclusively demonstrated. Found widely, including in Archean fossils~\cite{Sumner1997}, Antarctic lakes~\cite{Mackey2017} and hot springs~\cite{Castenholz1968}, these patterns can template more complex 3D morphogenesis~\cite{Shepard2010UndirectedMats,Mackey2017}.  They also provide rigidity~\cite{Shepard2010UndirectedMats} and enable collective mechanical responses, like rapid shape changes, to external cues~\cite{Castenholz1968,Pfreundt2023}.

Despite their importance to the development of complex life, and for \textit{e.g.} carbon-neutral biofuels~\cite{Farrokh2019}, no general mechanism has been identified to rationalize the collective behavior of filamentous cyanobacteria.  Here, we demonstrate that the emergent patterns of their colonies can be apprehended as the collective result of independently moving actors with simple interactions.  Distinctive features of filamentous cyanobacteria, such as their large aspect ratio and the tendency of a filament to follow the trail laid down by its head, enable the accurate prediction of the critical density and emergent lengthscale associated with collective ordering.

We investigate \emph{Oscillatoria lutea}, a typical strain of filamentous cyanobacteria, consisting of simple (non-branching, non-heterocystous) chains of cells.  Cultivation and measurement methods are provided as supplemental materials/appendicies.  In our cultures, the filaments have well-defined widths $\sigma = 4.2 \pm 0.2 \, \mu$m \cite{Faluweki2022} and lengths $L = 1.5 \pm 0.5$\,mm. In all cases here, error ranges report standard deviations. 

\begin{figure}
    \centering
    \includegraphics[width=\linewidth]{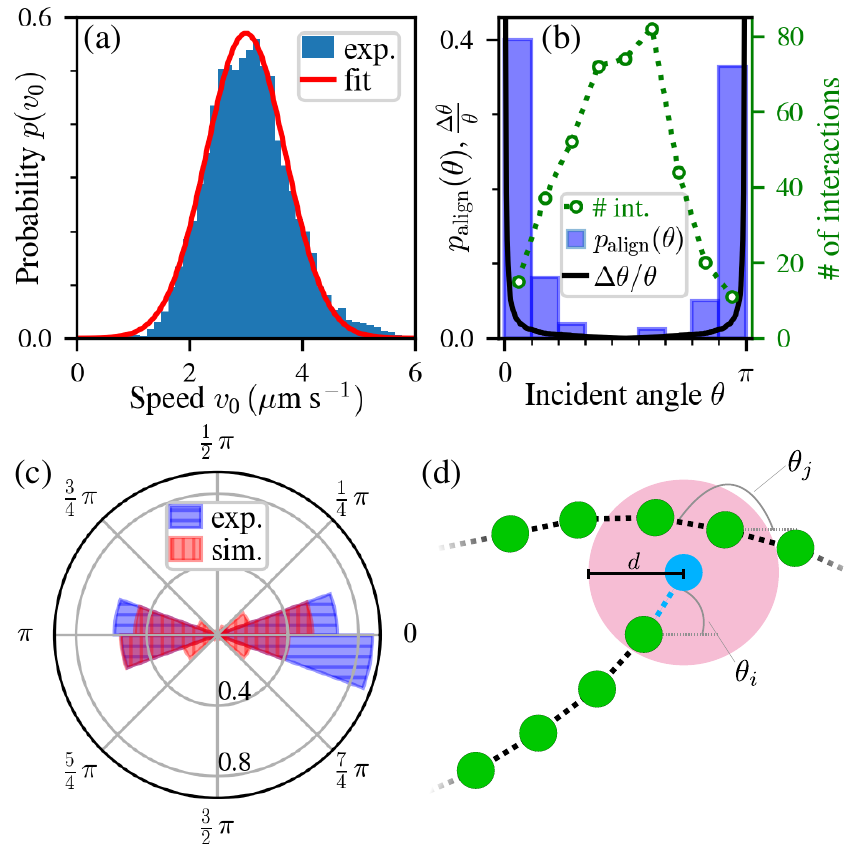}
    \caption{Filament behavior.  (a) The distribution of experimentally observed gliding speeds (blue) is well-fit by a Gaussian (red, used for simulations).  
    (b) {Histogram showing how the alignment probability (left axis, blue bars) and interaction frequency (right axis, green circles) depend on incidence angle $\theta$.  The data are experimental; see supplemental materials, for more details.} 
    In the model, an incident chain is deflected on average by the relative angle $\Delta\theta/\theta$ (black line).  
    (c) In bundles, the directions of motion have a nematic distribution: a polar histogram compares experimental (blue) and simulated (red) cases.  (d) Schematic of modeled interaction: when a chain's head is within distance $d$ of another chain, it experiences an aligning effect.
    }
    \label{fig:model_parameter_demo}
\end{figure}

In relative isolation, with area densities $\rho \simeq 1$\,mm$^{-2}$, filaments move at speeds $v_0 = 3.0 \pm 0.7~\mu$m\,s$^{-1}$, as shown in Fig.~\ref{fig:model_parameter_demo}(a).  They glide along smoothly curving paths, which we characterized by tracking the orientation $\theta$ of the tangent to each filament's midpoint through time.  The curvature $\kappa = d\theta/ds$ of the path $s(t)$ traced by any filament fluctuates slowly; the auto-correlation of $\kappa$ is well-described by an exponential relaxation with autocorrelation time $\tau = 470\pm 290$\,s (Fig.~S1). Isolated filaments are biased towards clockwise motion, as in related species~\cite{halfen1970,hoiczyk2000,Faluweki2022}.  However, from densities as low as $\rho=6$\,mm$^{-2}$ and up to $49$\,mm$^{-2}$ filaments adopt straighter shapes on average (Fig.~S2). These distributions of curvatures peak around zero, with standard deviation $\delta\kappa = 340 \pm 40$\,m$^{-1}$. 

To quantify the interactions between filaments, we observe cases where the head (leading end) of one filament approaches and intersects another filament.  In most such pairwise interactions there is no direct effect, rather the filaments simply pass over or under each other without changing paths.  However, about $4\%$ of the time the incident filament is deflected, turning to travel alongside the other filament, which typically remains unperturbed.  Aligning interactions only happen for small angles of incidence (Fig.~\ref{fig:model_parameter_demo}(b)), and result in the two filaments moving parallel or anti-parallel, depending on the angle of approach. After aligning, the filaments track each other for some distance, on average $430\,\mu$m, before one splits away. These interactions are fundamentally non-reciprocal~\cite{fruchart2021}, as the alignment response is path-dependent~\cite{scheibner2020}. 

The pairwise interactions promote the formation of bundles of aligned filaments, which can organize denser colonies into a higher-level architecture (Fig.~\ref{fig:motivation_fig}).  We confirmed the local nematic nature of this ordering by observing the motion of nearby filaments along one bundle, as in Fig.~\ref{fig:motivation_fig}(c).  All filaments in the bundle are well-aligned, with approximately equal fractions (223 versus 282 filaments; Fig.~\ref{fig:model_parameter_demo}(c)) traveling in either direction.  Between the bundles is a dilute `gas' of more randomly oriented filaments, similar in appearance to disordered colonies at lower densities. 

Some of these behaviors, such as nematic alignment and the tendency to form dynamic bundles and networks, are reminiscent of those of microtubules at an interface~\cite{sumino2012,doostmohammadi2018,sanchez2012}.  However, there are also conspicuous differences.  Critically, the average filament length is comparable to other characteristic lengths of this system, such as the filament's radius of curvature, or the emergent pattern lengthscale.   Hence, there is no \emph{a priori} clear separation of scales, and we will show that the elongated nature of the cyanobacteria filaments affects the nature of their collective self-organization.

A benefit of this perspective is that it leads directly to a relatively simple model that can be informed in all its parameter choices by experimental observations.  We treat the cyanobacteria as motile one-dimensional chains of point-like beads (Fig.~\ref{fig:model_parameter_demo}(d)), as befits their large aspect ratio, $L/\sigma> 100$.  For simplicity, all chains have length $L=1.5$~mm, and representative disorder is introduced \textit{via} their motion.  Their speeds are constant in time, but drawn from a normal distribution with average $v_{0}  = 3\,\mu$m\,s$^{-1}$ and standard deviation $0.7\,\mu$m\,s$^{-1}$, matching experimental values (Fig.~\ref{fig:model_parameter_demo}(a)).  The position $\bm{r}_{i,\alpha}$ of bead $\alpha$ of chain $i$ follows the track laid out by its head, so that $\bm{r}_{i,\alpha}(t)=\bm{r}_{i,\alpha-1}(t-\Delta t)$. At each time step, of duration $\Delta t$ the end bead is removed from the tail of each chain, and a new bead is added at its head, displaced by distance $v_i\Delta t$ at angle $\theta_i$ (Fig.~S4).
Similar models have been applied to isolated filaments \cite{du2022} and filaments on lattices \cite{schaller2010}.  Although this system has some similarities to active polymers~\cite{isele2015,duman2018,bianco2018,anand2018,kurzthaler2021}, those lack a unique curvature autocorrelation time, as each polymer segment fluctuates independently; in contrast, the fluctuations and curvature of our chains are solely determined by their heads.

Motivated by models of active nematic particles used to simulate microtubules \cite{nagai2015,sumino2012}, 
\emph{C. elegans} \cite{sugi2019} and \emph{Pseudanabaena} sp. \cite{yamamoto2021}, we now introduce a model of interacting active chains, appropriate to the behavior of filamentous cyanobacteria.  Here, the orientation $\theta_i$ and angular velocity $\omega_i$ of the head of each chain $i$ evolves by a modified Ornstein--Uhlenbeck process
\begin{align}
    \frac{d \omega_i}{d t} &= -\frac{1}{\tau} [\omega_i -J \mathcal{F}(\theta_i)] + \sqrt{2D_\omega}\xi_i(t) \label{eq:dt_omega},\\
    \frac{d \theta_i}{d t} &= \omega_i - J \mathcal{F}(\theta_i), 
    \label{eq:dt_theta}
\end{align}
where $\tau$ is the curvature autocorrelation time,  $J$ is an interaction strength, $D_\omega$ is a diffusion coefficient, and $\xi_i(t)$ introduces Gaussian white noise with zero mean and unit variance. $D_\omega$ is not directly accessible experimentally, but is linked to other parameters.  Without any filament-filament interactions, Eq.~\eqref{eq:dt_omega} produces a normal distribution of angular velocities with zero mean and variance $\langle \omega^2\rangle = D_\omega \tau$.  For chains with speed $v_0$ this translates into a curvature distribution with standard deviation $\delta \kappa = {\sqrt{ \langle \omega^2\rangle}}/{v_{0}}$.  Hence, $D_\omega = (v_0 \delta \kappa)^2/\tau$.  Finally, the interactions are modeled by $\mathcal{F}(\theta_i)=\frac{1}{N_{ij}} \sum_{j\sim i}\frac{\partial}{\partial \theta_i} U(\theta_i,\theta_j)$, using a nematic Lebwohl--Lasher potential, $U=-\cos\left[2(\theta_i-\theta_j)\right]$, averaged over the $N_{ij}$ chains within an interaction range $d$ of the head of chain $i$, where $\theta_j$ is the orientation of the nearest bead on chain $j$ (see Fig.~\ref{fig:model_parameter_demo}(d)).

\begin{figure}
    \centering
    \includegraphics[width=1\linewidth]{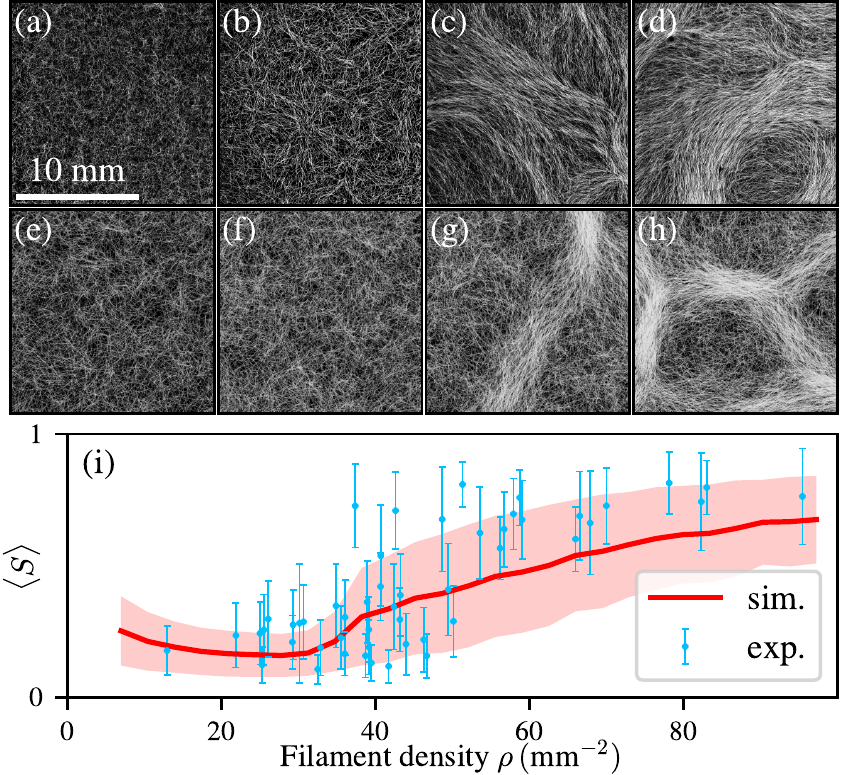}
    \caption{Collective behavior and order-disorder transition.  Panels (a--d) show micrographs of colonies at densities $\rho = 25, 31, 42$, and $59$\,mm$^{-2}$, respectively. Panels (e--h) show snapshots of simulations at comparable densities of $\rho = 24, 31, 41$ and $59$\,mm$^{-2}$. To avoid boundary effects, the simulated domains had sides $4.5\times$ larger than shown; panels are cropped to match the micrograph size.  (i) Order parameter $\langle S \rangle$, averaged over 1\,mm$^2$ blocks covering the experimental (blue) or simulated (red) domain; {see supplemental Fig.~S3 for more details.}  Error bars and shading give the standard deviation of $\langle S \rangle$ over the blocks.  At low $\rho$ the filaments are randomly aligned, but locally-nematic bundles and a reticulated structure emerge above $\rho\sim 40$\,mm$^{-2}$. 
    }
    \label{fig:order_density}
\end{figure}

\begin{figure*}
    \centering
    \includegraphics[width=1.\textwidth]{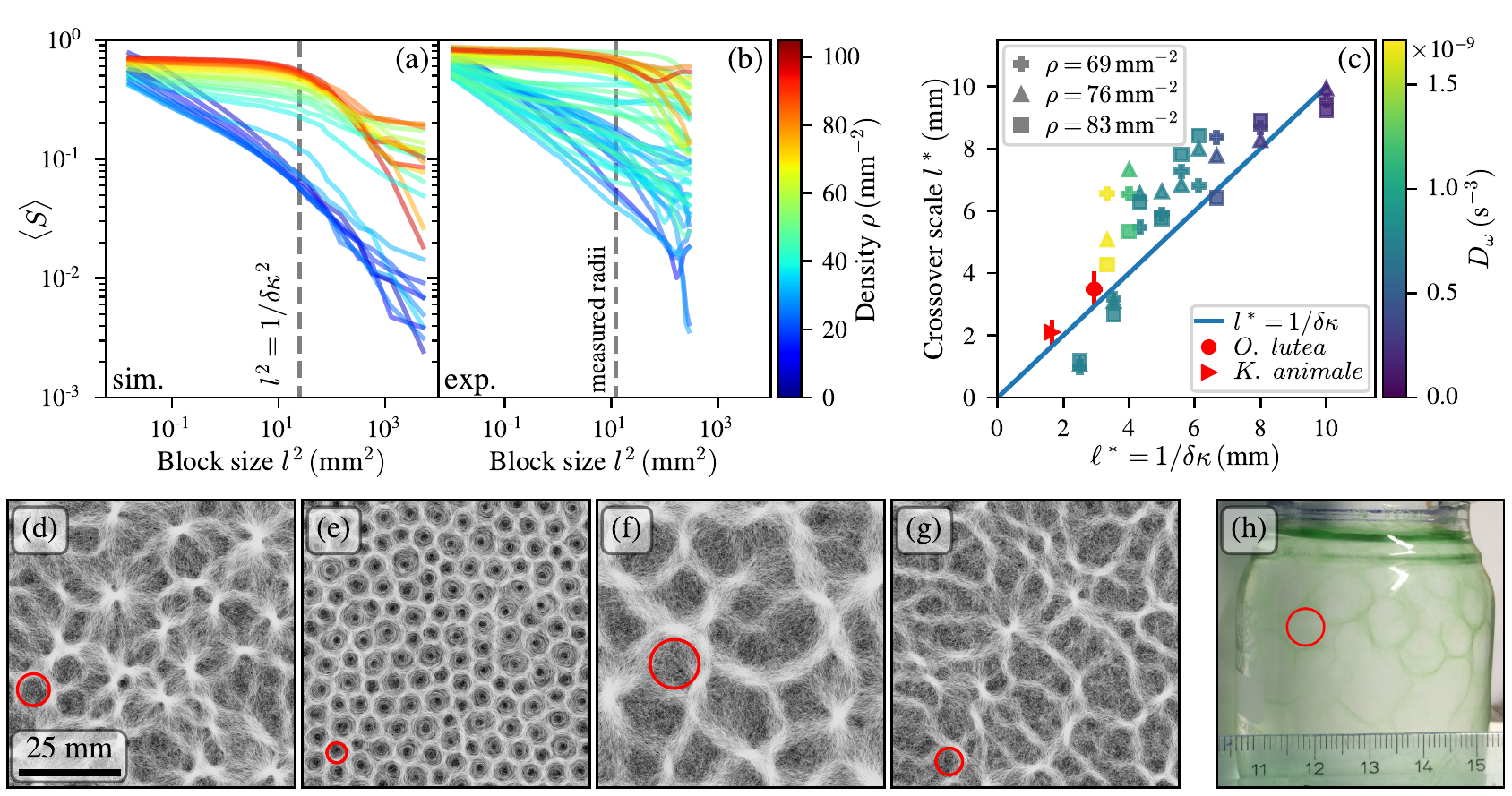}
    \caption{Emergence of large-scale patterning.  Finite-size scaling of the block-average order parameter $\langle S\rangle$ was investigated in (a) simulations and (b) experiments, by varying the block size $l$ for the same data shown in Fig.~\ref{fig:order_density}.   The power-law decay at low density indicates a disordered, isotropic state.  The emergence of structures at high density is marked by a plateau lasting until $l$ reaches the size of the emerging structures, which we term the crossover lengthscale $l^*$, after which a more rapid decay is observed.  (c) For different model parameters $l^*$ can be compared to the characteristic scale at which activity and fluctuations balance, $\ell^*$.  Snapshots show the resulting patterns for some simulations with 
    (d): $\rho=83\,$mm$^{-2}$, 
         $D_\omega =1.2\times 10^{-9}\,$s$^{-3}$, 
         $\tau=480\,$s.
    (e): $\rho=76\,$mm$^{-2}$, 
         $D_\omega =7.5\times 10^{-10}\,$s$^{-3}$, 
         $\tau=1920\,$s.
    (f): $\rho=83\,$mm$^{-2}$, 
         $D_\omega =7.5\times 10^{-10}\,$s$^{-3}$, 
         $\tau=320\,$s.
    (g): $\rho=69\,$mm$^{-2}$, 
         $D_\omega =1.7\times 10^{-9}\,$s$^{-3}$, 
         $\tau=480\,$s,
         and (h) for filaments growing naturally under typical incubation conditions.
    The characteristic scales of the patterns are shown by red circles of radius $\ell^*$. {The scale bar in (d) also applies to (e)-(g).}
    }
    \label{fig:order_gnf}
\end{figure*}

The model parameters were matched to experimental values of relatively isolated cyanobacteria, and fine-tuned based on the collective behavior at higher densities.  Unless otherwise stated, we set $\tau = 480\,$s, $\delta\kappa = 200\,$m$^{-1}$ 
(giving $D_\omega =7.5\times 10^{-10}\,$s$^{-3}$) and $d=5\,\mu$m, close to the observed values of 470 s, 340\,m$^{-1}$ and the filament diameter of 4.2\,$\mu$m, respectively.  The interaction strength, $J = 0.006\,$s$^{-1}$, was chosen by considering filaments meeting at an angle $\theta$.   On average, the effects of interactions are comparable if the incident filament is either deflected by a relative angle $\Delta \theta / \theta$, or by the whole angle $\theta$ with probability $p_\textrm{align}(\theta)$.  As shown in Fig.~\ref{fig:model_parameter_demo}(b), in this sense $J$ gives a similar average response to the observed interactions.    

Experimentally, colonies of cyanobacteria filaments are disordered at low density, but show emergent patterns at higher densities, Fig.~\ref{fig:order_density}(a--d). The simulated chains order in a similar way, Fig.~\ref{fig:order_density}(e--h), with reticulated structures appearing at higher $\rho$. Once formed, these structures remain relatively static. We quantify local order in the steady-state by the 2D nematic order parameter~\cite{Rezakhaniha2012,Persson2017,nishiguchi2017,Jordens}.  For this, each experimental or simulated system is divided up into blocks of size $l = 1$\,mm. At this scale the filament density is relatively homogeneous, but the blocks are large enough to have good statistics. The local order parameter $S = \langle\cos(2\hat{\theta})\rangle$ is measured for filament orientations $\hat{\theta}$ taken with respect to the local nematic director (see supplemental materials).  We then calculate $\langle S\rangle$ as a block average, which can quantify the emergence of local order, even in a globally heterogeneous system~\cite{rovere1990,nishiguchi2017}.

Both experiments and simulations show low nematic order at low densities.  At higher $\rho$, the appearance of collective structures is captured by a sharp increase in $\langle S \rangle$, as shown in Fig.~\ref{fig:order_density}(i).  Experimentally, the transition from a disordered state, with $\langle S\rangle \simeq 0.2$, to an ordered state of $\langle S\rangle \simeq 0.7$ is seen at a critical density of $\rho =$ 40--50\,mm$^{-2}$.  Simulations show a similar response,  and demonstrate that density inhomogeneities are correlated with the nematic ordering (Fig. S6). In no case is there any clear laning of filaments (see Figs.~\ref{fig:motivation_fig}(c), \ref{fig:model_parameter_demo}(c)), in contrast to stiff active rods~\cite{Shi2018,bar2020}. Varying the model parameters somewhat does not change the qualitative nature of the ordering transition, but does affect the critical value of $\rho$.  This quantifies prior speculation of a density-driven ordering transition~\cite{Shepard2010UndirectedMats,tamulonis2014}, and enables predictions.

For a `gas' of weakly-interacting filaments, can we predict when interactions will become important enough to lead to collective behavior?  For simplicity, consider filaments of density $\rho$ and speed $v_0$.  Filaments interact when they first cross, at some local tangent angle $\theta$ anywhere along a length $L$.  Averaging over all configurations, filaments thus present a mean cross-sectional length $\bar{L} = \langle L \sin{\theta}\rangle = 2L/\pi$ to each other.  As one filament advances, it then encounters others on average at frequency $f = \bar{L}\rho v_0 = 2L\rho v_0/\pi$.  Experimentally, only a small fraction $a$ of interactions cause alignment, so the rate of filament ordering scales as $af$.  Aligned filaments can also split up, which we assume happens randomly at rate $b$.  Under these representative assumptions, interactions should become important when the rates of filament alignment and breakup balance, $a f \simeq b$, and this cross-over condition defines a characteristic density $ \rho_{c}={ \pi b}/({2a Lv_0})$.  Using experimental values, $a=0.04$ and $b=0.007$\,s$^{-1}$ (see supplemental materials), predicts $\rho_{c} \sim O(50)$\,mm$^{-2}$. A disordered gas of filaments would be expected for densities $\rho \ll \rho_{c}$, with ordered states starting to appear at densities $\rho \approx \rho_{c}$.  This prediction agrees well with the density of the order-disorder transition shown in Fig.~\ref{fig:order_density}.

We can rationalize the emergent lengthscale of the reticulate pattern as a signature of the balance struck between activity and fluctuations. By nondimensionalizing Eqs.~\eqref{eq:dt_omega}--\eqref{eq:dt_theta}, the ratio between the angular rate of change and diffusion defines a P\'eclet number, $\mathrm{Pe} = v_0/(\ell\sqrt{D_\omega\tau})$, where $\ell$ is some reference length.  In the steady-state, the nonequilibrium probability fluxes associated with active motion and curvature fluctuations will strike a balance, determining a specific lengthscale $\ell^*$, corresponding to $\mathrm{Pe} =1$, as the smallest scale over which patterns can emerge.  Using $D_\omega = v_0^2\delta\kappa^2/\tau$ we predict $\ell^* = v_0(D_\omega\tau)^{-1/2} = 1/\delta\kappa\approx 5$~mm.

To substantiate this prediction, we perform a scaling analysis~\cite{Persson2017,nishiguchi2017} of how the block-averaged order parameter $\langle S \rangle$ depends on the block size $l$.  Figure~\ref{fig:order_gnf}(a,b) shows the results for simulations and experiments. At low densities we see the power-law decay expected for a disordered system \cite{nishiguchi2017}. With increasing $\rho$, the experimental data is noisier, but potentially shows structure developing more continuously than in the simulations.  For $\rho>40\,$mm$^{-2}$, $\langle S(l) \rangle$ develops two distinct regimes: a plateau at low $l$, reflecting the local order within bundles, and a faster decay at large $l$.  From the position of the crossover between these responses we extract a lengthscale $l^*$ (Fig.~\ref{fig:order_gnf}(c), methods in supplemental materials).  The drop in $\langle S \rangle$ above $l^*$ is attributed to bundles with different orientations appearing within the same block.

In simulations, we explore the dependence of $l^*$ on the model parameters, by varying $\tau$, $D_\omega$, and $\rho$. Some steady-state snapshots are shown in Fig.~\ref{fig:order_gnf}(d-g).  While the fine details of the patterns vary, $l^*$ is always consistent with the radius of the emergent structures, with no significant dependence on $\rho$.  As shown in Fig.~\ref{fig:order_gnf}(c), this feature size generally matches the characteristic length $\ell^*=1/\delta\kappa$ predicted \textit{via} $\mathrm{Pe}$.  In \emph{O. lutea}, the radius of the structures of dense colonies is $l^*=3.5\,\pm\,0.6$\,mm, see Fig.~\ref{fig:order_gnf}(h), consistent with $\ell^* = 1/\delta\kappa = 2.9$\,mm.  Repeating measurements on the related species \emph{Kamptonema animale} (see supplemental materials; $l^*=2.1 \pm 0.4$ mm, $\ell^* = 1.7 \pm 0.1$ mm) further confirms this correspondence of length-scales.

Summarizing, we studied colonies of filamentous cyanobacteria and their collective organization. The filament length is comparable to other scales in this problem (\emph{e.g.} curvature) and can couple with them; one cannot assume separation of scales. A nonequilibrium theoretical model accounting for fluctuations, large aspect ratios, motility, and nematic alignment reproduces the structure of reticulate patterns seen in the lab~\cite{Shepard2010UndirectedMats} and nature~\cite{Castenholz1968, Mackey2017, cuadrado2018}.  Our results thus point to a new class of active matter characterized by the following features: (\emph{i}) Elongated filaments with position-dependent orientation and multiple interaction sites along each filament. (\emph{ii}) Gliding motility induced by polar forces \cite{abbaspour2021}, unlike extensile/contractile microtubule-kinesin systems \cite{sanchez2012}; and (\emph{iii}) path-tracking dynamics of the body following its head, subject to fluctuations and active motion, which are ultimately responsible for the reticulate pattern and lengthscale selection.  Cyanobacteria are an important class of microbial life, and among the earliest form of multicellular organisms. We note that the parameters governing their self-organization identified here are evolutionarily selectable traits, influencing collective responses~\cite{Castenholz1968,Pfreundt2023}, mechanical properties~\cite{Shepard2010UndirectedMats} and 3D morphologies~\cite{Shepard2010UndirectedMats,Mackey2017}, and can inform the study of the fossil record~\cite{Sumner1997,cuadrado2018}. 

\acknowledgements
We thank Maike Lorenz (SAG G\"ottingen) for support with cyanobacteria cultures, Stefan Karpitschka (MPIDS) and Jack Paget (Loughborough) for discussions and Graham J. Hickman (NTU) for microscopy support.  Microscopy facilities were provided by the Imaging Suite at the School of Science and Technology at Nottingham Trent University. Numerical calculations were performed using the Sulis Tier 2 HPC Platform funded by EPSRC Grant EP/T022108/1 and the HPC Midlands+ consortium. We gratefully acknowledge use of the Lovelace HPC service at Loughborough University.  M.K.F. was partly sponsored by the Malawi University of Science and Technology. This work was supported by the Max Planck Institute for Dynamics and Self-Organization (MPIDS)

\section*{Supplemental Information}

\appendix
\renewcommand{\thefigure}{S\arabic{figure}} 
\setcounter{figure}{0} 

\begin{figure}
    \centering
    \includegraphics[width=1.\linewidth]{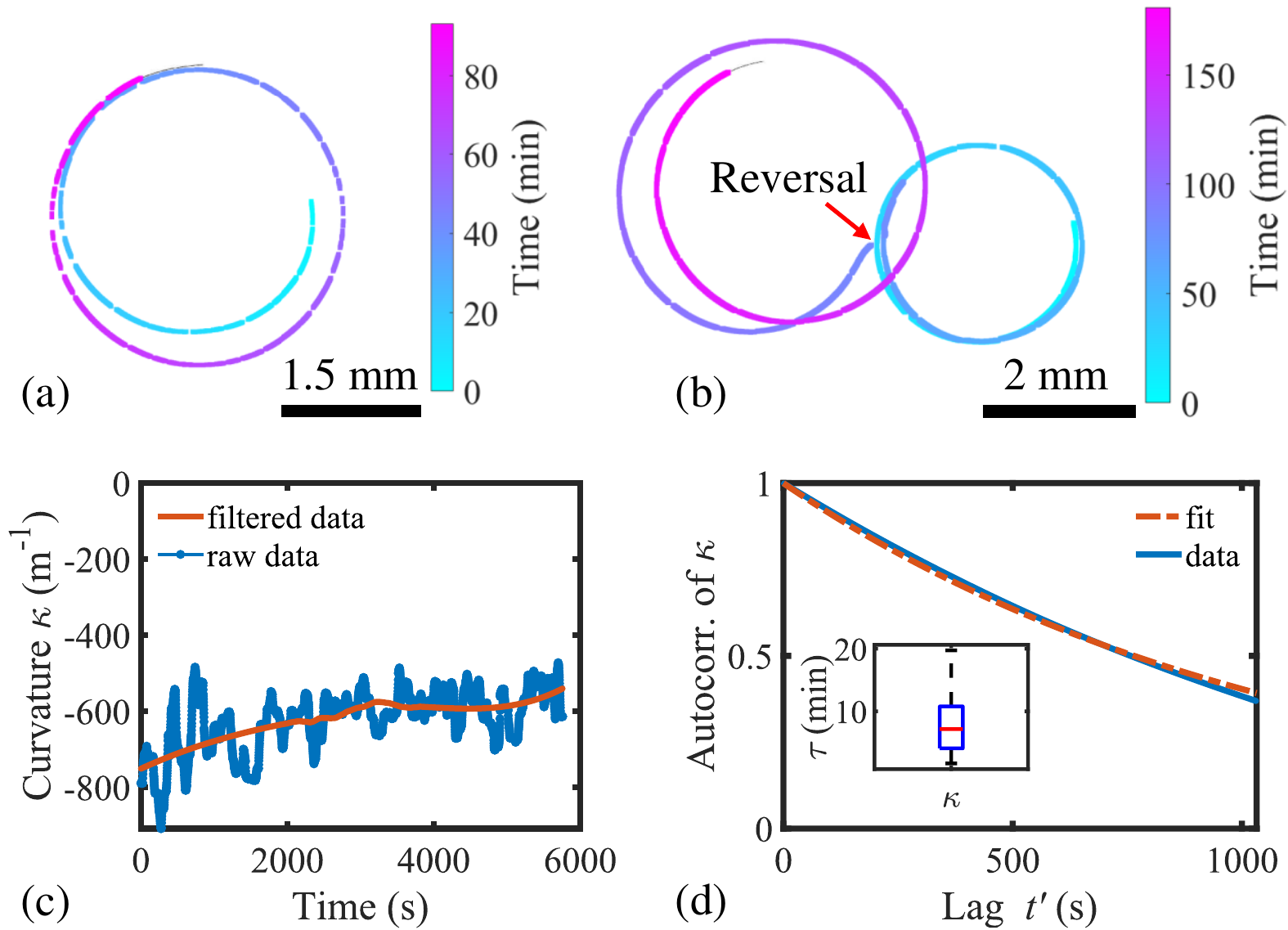}
    \caption{Motion of isolated \emph{O. lutea}. (a) Filaments followed smoothly curving paths, as shown here by the track of one midpoint over time.  (b) The filaments occasionally reversed their direction of motion, but their path curvature was maintained across such events, as in the track shown here.  (c) The path curvature, $\kappa$, along any track fluctuated in time, with negative values indicating clockwise motion; data here are from the track in panel (a), before and after smoothing. (d) The autocorrelation of the filtered data is well fit by an exponential decay with a correlation time $\tau$. The insert shows the distribution of $\tau$ for different filaments.}
      \label{fig:parameter_motion}
\end{figure}

\begin{figure*}[th]
    \centering
    \includegraphics[width=1.\linewidth]{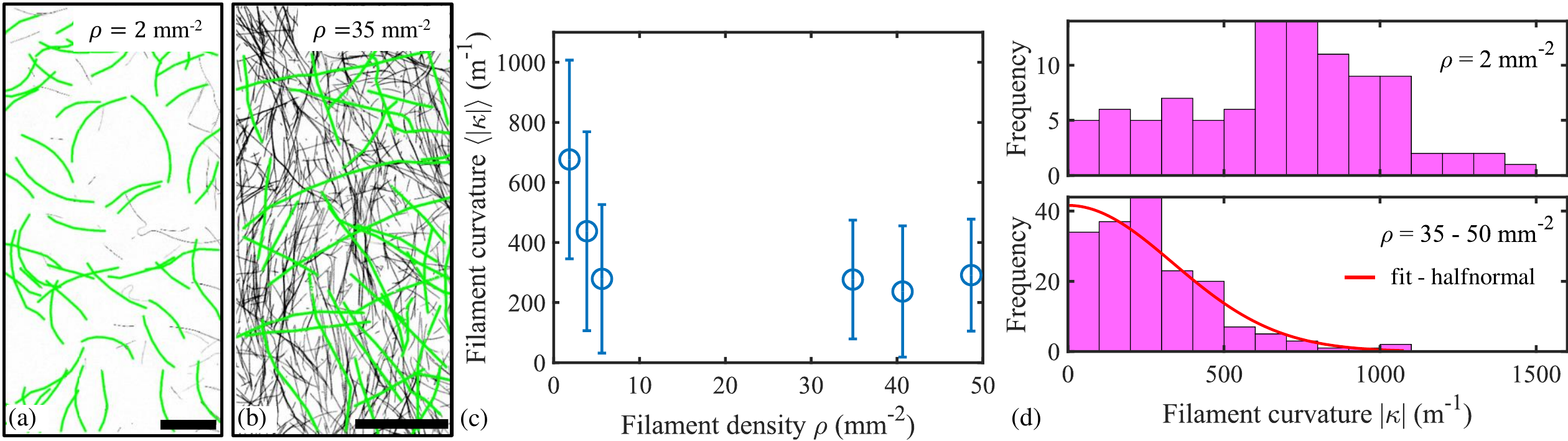}
    \caption{Filament curvature changes with density $\rho$. (a) At low densities filaments are visibly curved in shape while (b) at higher densities the filaments are straighter; scale bars are 1 mm.  Filaments highlighted in green have been manually masked for curvature measurements. (c) The average curvature drops with $\rho$, while the standard deviation (error bars) remains relatively constant. (d) Histograms of $|\kappa|$ show the bias towards a preferred curvature at low $\rho$.  At higher $\rho$ the curvature distribution is consistent with a normal distribution centered around zero curvature.}
      \label{fig:parameter_curvatures}
\end{figure*}

\section*{Experimental methods}
\textbf{Culture preparation.} Stock of \emph{Oscillatoria lutea} (SAG 1459-3) was maintained in a medium of BG11 broth (Sigma-Aldrich) diluted to a ratio of 1:100 with deionized water. Following Lorentz \textit{et al.}~\cite{Lorenz2005PerpetualCultures}, samples were incubated at $20 \pm 1\,^{\circ}$C, with warm-white LEDs (color temperature of 2800\,K) providing a photon flux of $10 \pm 2$\,$\mu$mol m$^{-2}$\,s$^{-1}$ on a 16\,h day + 8\,h night cycle. For sampling, material was transferred into a 100 ml bottle half-filled with medium and shaken mildly to separate the filaments. Samples were then drawn into a syringe and added dropwise to 6-well plates (34\,mm well diameter) three-quarters filled with medium. The colony density $\rho$ was controlled by varying the number of drops added to each well. The well plates were covered and left in the incubator for 72 hours before imaging. 

\textbf{Imaging.} We used a confocal laser scanning microscope (Leica TCS SP5) in bright field and fluorescence modes. Fluorescence of the chlorophyll-a in the cyanobacteria was excited by the 514 nm line of the argon laser at 29\% power. The light emitted was detected through a 620--780 nm band-pass filter by a HyD hybrid detector at 100\% gain. Images were observed with a PL Fluotar 10X/0.3na air objective with a pinhole of 70.8 $\mu$m. Scanned image frames were 512$\times$512 pixels (1.55$\times$1.55 mm$^2$) with a scan rate of 400 Hz. No averaging or integration was applied during collection. Dynamic measurements were made at fixed positions or with manual tracking. Wide area imaging (\textit{e.g.} Figs.~1, 3) used the microscope's tile scan protocol, and were reconstructed from overlapping frames collected by the rapid progressive scan of regions of interest.  Images were binarized in Matlab using the \texttt{adaptthresh} algorithm to ensure a consistent appearance of individual filaments, then skeletonized and despurred for further analysis.

\section*{Parameter measurement}

\begin{figure*}
 \begin{center}
  \includegraphics[width=1\linewidth]{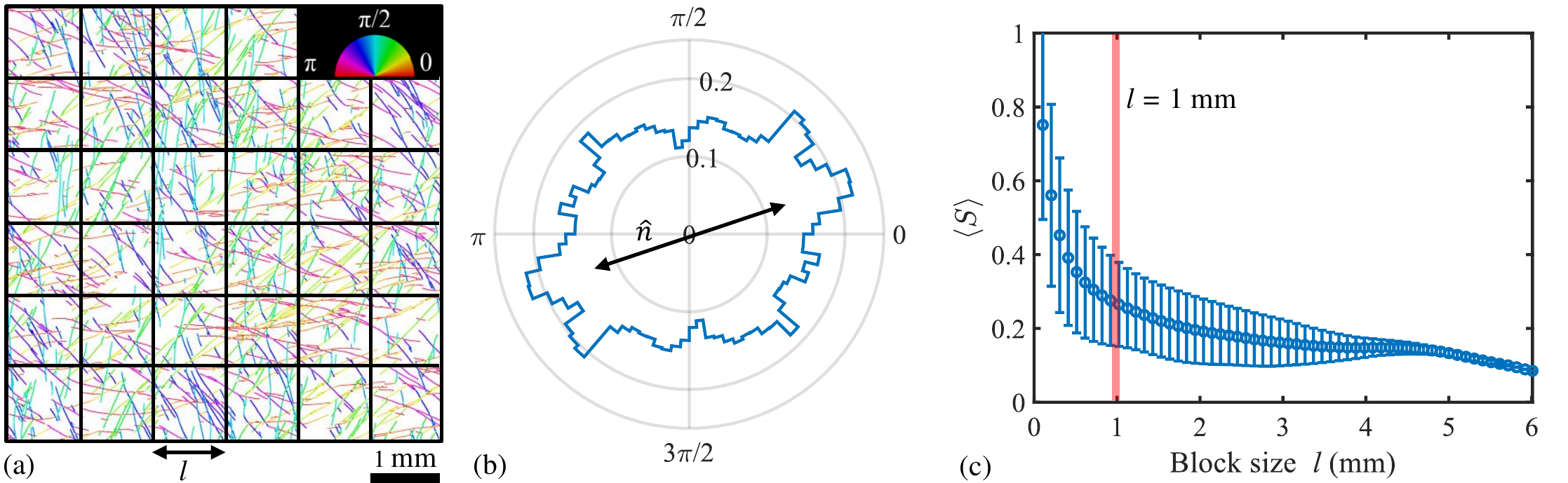}
  \caption{Calculation of the experimental order parameter. (a) Images were skeletonized, with an  orientation or direction assigned to every pixel on the skeleton, and partitioned into blocks of size $l$; the legend shows the color-coding of the orientation.  (b) Locally, $S$ was calculated from the distribution of orientations within any particular block, with respect to the local director, $\hat n$.  (c) A block average then gives the global order parameter, $\langle S \rangle$, which depends on the size of the blocks used, $l$. Error bars give the standard deviation across blocks.  For the analysis in Fig.~3 we use a representative block size of $l = 1$\,mm (red line).}
  \label{fig:parameter_box}
 \end{center}
\end{figure*}

\textbf{Isolated filaments.}  The filaments had lengths of $L = 1.5 \pm 0.5$\,mm, as measured along their skeleton, and cross-sectional diameters of $\sigma = 4.2 \pm 0.2\,\mu$m, as measured in Ref.~\cite{Faluweki2022}.  We followed the motion of 23 isolated ($\rho < 1$\,mm$^{-2}$) filaments over time, generating time series of the positions of the head, tail and midpoint of each skeletonized filament over  observation periods of up to 3 hours.   The distribution of instantaneous speeds, $v_0$, in Fig.~2(a) was measured using the midpoint tracks and a 40-point moving window.  There was no significant correlation of speed with filament length.

Figure~\ref{fig:parameter_motion}(a) shows the isolated filaments tracing smoothly curving paths, along which the curvature fluctuates over time.  Filamentous cyanobacteria can intermittently reverse the direction of their motion \cite{gabai1985,tamulonis2014}, see Fig.~\ref{fig:parameter_motion}(b).  Such reversals can affect the collective behavior of filaments that are sufficiently confined so as to prevent filament crossings~\cite{abbaspour2021}, although this limit is far from our experimental conditions.   We observed 25 spontaneous reversals during 23 h of single-filament tracking, at intervals between 10 minutes and several hours.  After a reversal, a filament typically continued along a new path with a curvature close to its pre-reversal value.   

Path curvatures, $\kappa = d\theta/ds$, were quantified from the time-lapse image sets.   A tangent-line fit to the central half of each filament was used to measure its orientation $\theta$, and the path coordinate $s$ was taken from the track of its midpoint.  A numerical derivative for $d\theta/ds$ was then calculated using a 40-point moving window in time.  To remove high-frequency noise, resulting from the numerical differentiation, the curvature data was smoothed by a third-order Savitzky-Golay filter (method adapted from~\cite{sumino2012}, demonstrated in Fig.~\ref{fig:parameter_motion}(c)).  The smoothed path curvature data are consistent with time series of curvatures measured by fitting circular arcs to the filament skeletons (methods adapted from~\cite{Faluweki2022}).  The path curvature autocorrelation function was calculated as $\langle\kappa(t)\kappa(t+t^\prime)\rangle / \langle\kappa^2\rangle$, for delay $t^\prime$.  For each of the 23 filament tracks a correlation time $\tau$ was found by fitting the exponential relaxation $e^{-t^\prime/\tau}$ to the autocorrelation function, see Fig.~\ref{fig:parameter_motion}(d).  The distribution of $\tau$ for all filaments studied had a mean of 470 s and standard deviation of 290 s.  To check the robustness of these methods, correlation times of $540 \pm 300$\,s were calculated in the same way, but starting from the time series of filament curvatures (\textit{i.e.} as fit by circular arcs).  

\textbf{Interacting filaments.}  We quantified the pair-wise interactions of gliding filaments in colonies with intermediate densities of $\rho \simeq 10$\,mm$^{-2}$. Filaments interacted through contact, when their paths crossed.  400 such interactions were tracked; in each case the angle of incidence was taken with respect to the forward directions of motion of the filaments at the point of contact.   As summarized in Fig.~2(b), the incident filament either turned to follow beside the filament it met, or the two filaments crossed over/under each other without altering their paths.  Only 16 events resulted in alignment, giving this outcome a relative probability of $a = 0.04$.  Of these, there were 10 cases of parallel alignment, and 6 of anti-parallel alignment, where the newly bundled filaments moved in opposite directions along adjacent paths.   

After aligning, filaments travel together for some time, before one filament breaks off onto a separate path.  After each of the 16 alignment events, we tracked the distance traveled before the pair broke up.  The mean distance traveled while being aligned was $d_b = 430\,\mu$m with standard deviation 200 $\mu$m.  A similar effect was seen at walls, where filaments that hit a wall curved to follow it for an average of $520\,\pm\,280\,\mu$m, before breaking away (averaged over 100 observations).  A representative rate of filament breakup, $b = 0.007$\,s$^{-1}$, was calculated as $v_0/d_b$.

The motion of filaments in bundles was characterized using time-lapse images from seven locations clustered along a single long bundle..  Directions of the motion of 505 filaments were measured by hand, in ImageJ~\cite{ImageJ}. For each location, we defined a local average orientation, maintaining a consistent sense of the motion along the bundle (in this case, with angles near zero implying motion roughly from the top to the bottom of the image).  Figure~2(c) shows the relative directions of motion of the filaments, measured with respect to their local nematic director.

Finally, we observed the effects of interactions on filament shapes.  In isolation \textit{O. lutea} filaments tend to glide in clockwise rotation, with a preferred curvature of $540\,\pm\,230$\,m$^{-1}$~\cite{Faluweki2022}.  Here, we measured curvature by manually masking individual filaments in thresholded images of colonies at various densities, as shown in Fig.~\ref{fig:parameter_curvatures}(a,b).  The masked filaments were skeletonized, despurred, and circular arcs were fit to their shapes.  The mean and standard deviation of the resulting distributions of unsigned curvatures are shown in Fig.~\ref{fig:parameter_curvatures}(c).  At low densities,  Fig.~\ref{fig:parameter_curvatures}(d--top), the results are similar to isolated filaments.  As their density increases, Fig.~\ref{fig:parameter_curvatures}(d--bottom), the filaments become straighter on average, and the peak of the curvature distribution shifts towards zero.  As a best estimate of filament shapes in dense colonies, we combined observations from $\rho = 35$, 41 and 49\,mm$^{-2}$, and fit the results with a Gaussian distribution of zero mean.  Figure~\ref{fig:parameter_curvatures}(d) shows the fit, which gives a representative spread of curvatures of $\delta\kappa= 340\,\pm\,40\,$m$^{-1}$.

\begin{figure}[b]
    \centering
    \includegraphics[width=1.\linewidth]{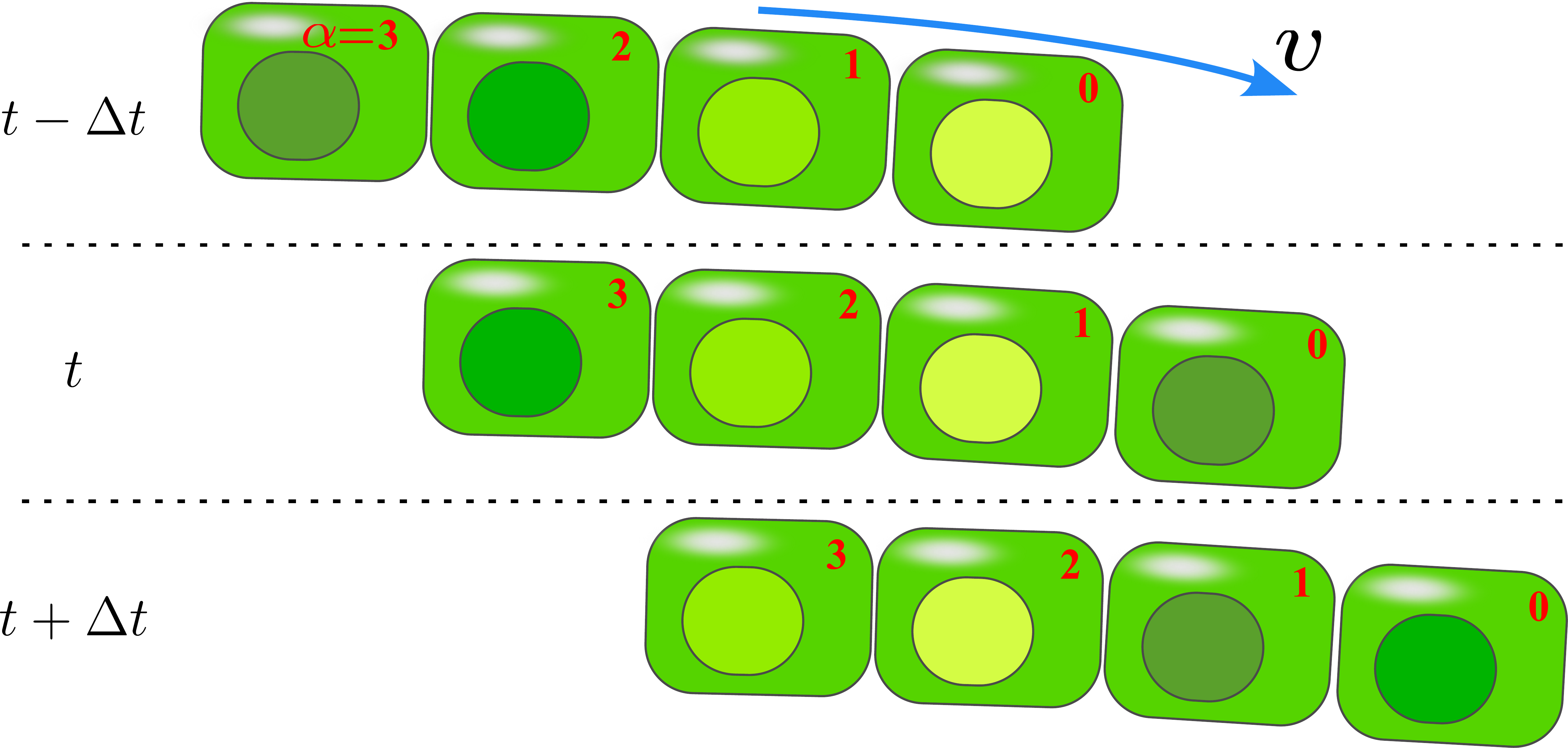}
    \caption{Sketch of the motion of modeled filaments. For simplicity, a filament of four beads (marked from $\alpha = 0$ to $3$, where $0$ is the head) is shown at three generic time steps: $t-\Delta t$, $t$, and $t+\Delta t$. At each time step, the tail bead ($\alpha = 3$) is removed and a new bead is placed in the front of the filament as the new head. An arrow indicates the direction of motion. 
    }
    \label{fig:SI_sketch_beads}
\end{figure}

\begin{figure*}
    \centering
    \includegraphics[width=\textwidth]{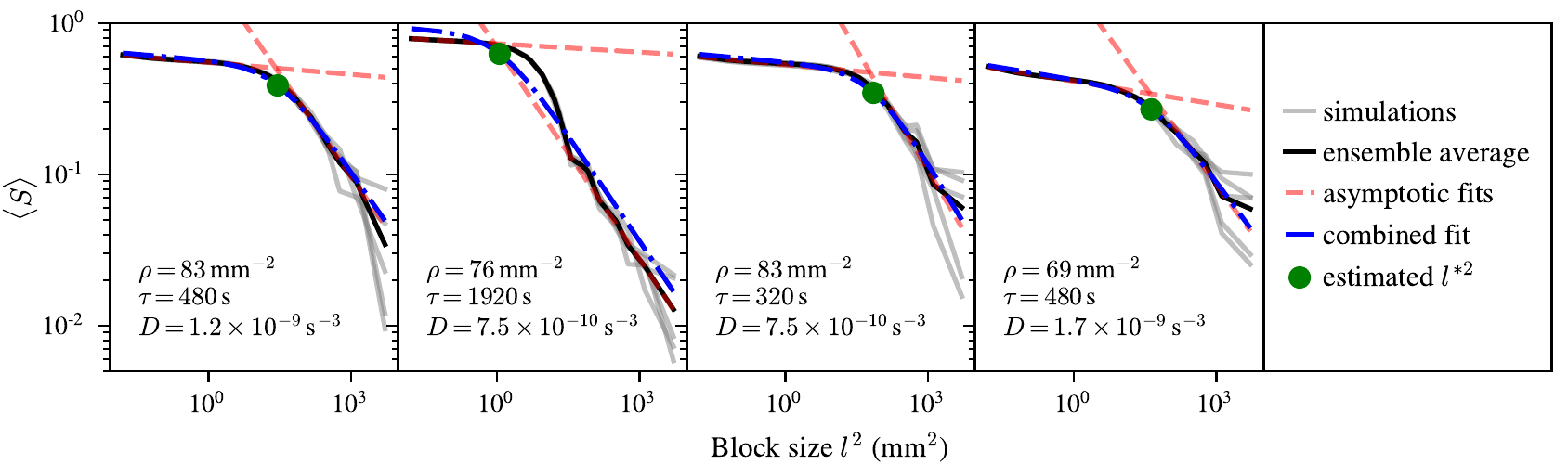}
    \caption{Examples of the procedure used to identify the scale of reticulate patterns. The order parameter $\langle S \rangle$ is calculated within blocks of various size $l$, and averaged (black line) over five independent simulations (gray lines) for each set of parameters $\rho,\tau,D_\omega$.  Linear fits identify the asymptotic slopes (red dashed lines), which are smoothly connected to estimate a crossover length scale $l^{*}$ (green circle). The four examples shown correspond to the simulations in Fig.~4(d--g).}
    \label{fig:SI_length_scales}
\end{figure*}

\textbf{Order parameter.}
The 2D nematic order parameter, $\langle S \rangle$, was calculated using the GTFiber App~\cite{Persson2017} \textit{via} the structure tensor method~\cite{Persson2017,nishiguchi2017}.  This assigns an orientation from 0 to $\pi$ to each pixel on an image skeleton, and then divides the image up into blocks of size $l$, as in Fig.~\ref{fig:parameter_box}(a).  In each block the local order parameter $S = \langle\cos(2\theta_n)\rangle$ is calculated, where the orientation $\theta_n$ is measured with respect to the local director, or average orientation of pixels, within that block (see Fig.~\ref{fig:parameter_box}(b)).   The global order parameter $\langle S \rangle$ is taken as the average of $S$ over all non-empty blocks.  As shown in Fig.~\ref{fig:parameter_box}(c), its value depends on the block size, $l$.

We characterized $\langle S\rangle$ in 50 colonies prepared identically (see culture preparation), but with different $\rho$.  In each case a $17 \times 17$\,mm$^2$ region of interest was cropped from the center of a confocal image of the whole colony, to minimize the influence of the chamber boundaries.  The data in Fig.~3 use a representative block size of $l = 1$\,mm, which is large enough to provide a good statistical average within each box, but small enough to still give a homogeneous sampling.    Results in Fig.~4 are prepared in the same way, but with varying $l$.

\textbf{\textit{K. Animale.}} Key measurements were repeated for \emph{Kamptonema animale} (SAG 1459-6) under similar incubation and observation conditions as for \emph{O. lutea}.  Both species belong to order Oscillatoriales~\cite{Strunecky2014}, have structures of simple chains of cells, and show similar emergent patterns. The parameters measured were filament speed ($v=2.5\pm 0.1$ $\mu$m s$^{-1}$), autocorrelation time ($\tau=526\pm 5$ s), variation of filament curvature ($\delta\kappa=605 \pm$ 58 m$^{-1}$) and reticulate radius ($l = 2.1 \pm 0.4$ mm). 

\begin{figure}[b]
    \centering
    \includegraphics[width=1.\linewidth]{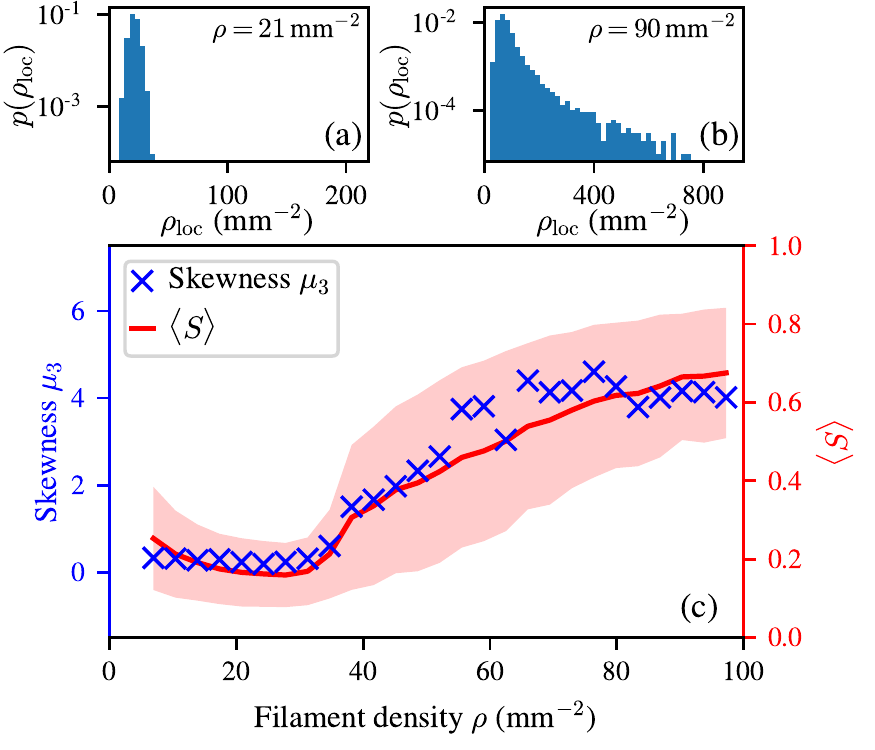}
    \caption{Characterization of density distributions.  (a) At low global density $\rho$, the distribution of local densities $\rho_{\mathrm{loc}}$ in $l = 1$ mm blocks is narrowly peaked around the mean.   (b) At higher densities, this distribution develops a long tail corresponding to dense bundles separated by sparser regions.  (c) The skewness $\mu_3$ of the density distributions begins to rise at a global density of $\rho\approx 40$ mm$^{-2}$, in line with the development of  the nematic order parameter $\langle S\rangle$.}
    \label{fig:skewness}
\end{figure}

\section*{Numerical Methods}

\textbf{Simulations.}  We used $72\times72$\,mm$^2$ domains with periodic boundary conditions to model the motion of $N=$ 36\,000--504\,000 filaments, corresponding to the density range $\rho =$ 7--97\,mm$^{-2}$.  Each filament $i$ had a fixed speed $v_i$ and was discretized into a chain of beads with positions $\bm{r}_{i,\alpha}$, where the index $\alpha$ counts beads from the head, $\alpha = 0$, to the tail.  At each time step, of size $\Delta t = 0.5$\,s, the bead at the tail end of each filament is removed, and a new bead is added as its new head.  All other beads increment their index, $\alpha \rightarrow \alpha + 1$, without changing position, as sketched in Fig.~\ref{fig:SI_sketch_beads}.   This economical move reproduces the experimental behavior where the head of a filament leads, while the rest of the filament follows in its track. 

To determine the updated location and orientation of the filament heads after each time step, we use the Euler--Maruyama algorithm to solve Eqs.~(1-2).  If the head of a filament is within the interaction range $d$ of any links between two beads of another filament, the interaction potential $U(\theta_i,\theta_j)$ is calculated based on the current orientation of the head, $\theta_i$, and the orientation of the closest link, $\theta_j$.   All measurements are made after an equilibration time of $10^5$\,s of simulated time. 

\textbf{Determining crossover length.}  The scaling of the block-average order parameter $\langle S \rangle$ with the block size $l$ can reveal structural information~\cite{Persson2017,nishiguchi2017}.  We calculated $\langle S \rangle$ for simulations in the same way as the experiments (see Fig.~\ref{fig:parameter_box}), based on the orientations of beads within each block, and averaging over an ensemble of five independent simulations with different random seeds.  At higher densities there were two distinct scaling regimes.  As shown in Fig.~\ref{fig:SI_length_scales}, there is a crossover lengthscale between these limits, $l^*$ (green dot), which we use as an indication of the size of the emergent structures.  We found that a robust way to identify $l^*$ was through the intercept of the two asymptotic power laws.  To this end, we performed linear least-squares fits of $\log\langle S \rangle=n_1\log(l^2)+b_1$ for $l^2<1\,$mm$^2$ and $\log\langle S \rangle=n_2\log(l^2)+b_2$ for $l^2>100\,$mm$^2$.  To smoothly connect the two cases we then fit 
\begin{align}\label{eq:smooth_connection}
    \log\langle S \rangle = (n_2-n_1)\log(l^2-l^{*2}) + n_1\log(l^2)+b \nonumber
\end{align}
to the whole range of data, with $l^*$ and $b$ as fitting parameters.  The resulting fits are shown in Fig.~\ref{fig:SI_length_scales} for some different parameter choices of $\rho$, $\tau$, and $D_\omega$.  As shown in Fig.~4(d--g), $l^*$ gives a good estimate of the average radius of the emergent structures of the reticulated patterns.

\textbf{Density distributions.} To quantify the degree of density inhomogeneity, we compute local-density histograms by dividing up the simulation domain into a regular grid of blocks of size $l=1\,$mm and finding the local density $\rho_\text{loc}$ in each grid block.  As the global filament density $\rho$ increases, these distributions evolve from a symmetric shape to an asymmetric distribution with a long tail that represents the condensed bundles, see Fig.~\ref{fig:skewness}(a,b).  To capture this change, we measure the third standardized moment as given by the skewness $\mu_3 = \langle (\rho_{loc}-\rho)^3/\sigma_\rho^3 \rangle$, where $\sigma_\rho$ is the standard deviation of the density distribution. The results are shown in Fig.~\ref{fig:skewness}(c). The skewness is a measure of the density inhomogeneity in the system and begins to rise at a global density of $\rho\approx 40$ mm$^{-2}$, which matches the value of the transition to a reticulate pattern as captured by the appearance of nematic order. 


%

\end{document}